\documentclass[twocolumn,english,aps,prb,floats,showpacs]{revtex4}
\usepackage[T1]{fontenc}
\usepackage[latin9]{inputenc}
\usepackage{amsmath}
\usepackage{amssymb}
\usepackage{graphicx}
\usepackage{esint}

\makeatletter

\newcommand{\bm}
\makeatother

\usepackage{babel}
\begin{document}

\title{Damping of a nanocantilever by paramagnetic spins}

\author{E. M. Chudnovsky and D. A. Garanin}

\affiliation{Physics Department, Lehman College, City University of New York \\
 250 Bedford Park Boulevard West, Bronx, New York 10468-1589, USA}

\date{\today}
\begin{abstract}
We compute damping of mechanical oscillations of a cantilever that contains flipping paramagnetic spins. This kind of damping is mandated by the dynamics of the total angular momentum, spin + mechanical. Rigorous expression for the damping rate is derived in terms of measurable parameters. The effect of spins on the quality factor of the cantilever can be significant in cantilevers of small length that have large concentration of paramagnetic spins of atomic and/or nuclear origin. 
\end{abstract}

%\pacs{62.25.-g, 76.60.Es, 07.10.Cm, 85.85.+j}

\maketitle

\section{Introduction}

\label{intro}

Small cantilevers have various applications in atomic force microscopy (AFM),  in micro- and nanoelectromechanical  systems (MEMS and NEMS), and for biological chemical detection \cite{Eom-PR2011}. Submicron cantilevers have recently permitted spatial resolution of the AFM that is sufficient to visualize tiny details of the DNA double helix near physiological conditions \cite{Leung-NL2012}. Further minituarization of cantilevers has potential to revolutionize technology and medicine. The accuracy of detectors based upon nanocantilevers relies on the quality factor of the cantilever, see, e.g., Ref.\ \onlinecite{Davis-APL2010}. Mechanical motion of the cantilever is related to the dynamics of the angular momentum. Coupling of the mechanical angular momentum and the angular momentum associated with the magnetic moment of a ferromagnetic body is described by Barnett and Einstein - de Haas effects \cite{Barnett,EdH}. In a paramagnetic body that coupling is less transparent. The question is whether thermal flipping of atomic and nuclear angular momenta inside a non-magnetic cantilever can affect its quality factor. For a relatively large cantilever, having high moment of inertia, it seems unlikely that tiny angular momenta of atoms and nuclei may have any significant effect on the cantilever. However, as we shall see, the effect scales inversely with the square of the length of the cantilever and it may become important for nanoscale cantilevers. 

Coupling of cantilevers to classical magnetic moments has been studied in the past in the context of the possibility to reverse the magnetic moment by mechanical motion \cite{Kovalev-PRL2005,Kovalev-PRB2007}. Einstein - de Haas effect in a magnetic cantilever has been measured \cite{Wallis-APL2006} and explained \cite{JCG-PRB2009} by the motion of a domain wall. Coupling of cantilevers to quantum spins has been investigated theoretically \cite{JC-PRL2009,Kovalev-PRL2011,GC-PRX2011}. Experiment has progressed to the measurement of a single molecular spin in a NEMS obtained by drafting of a single-molecule magnet on a carbon nanotube \cite{Wern-NataureNano2013,Wern-ASC-Nano2013}. Theory of such experiment that treats both the spin and the cantilever as quantum objects has been developed in Ref.\ \onlinecite{OCG-PRB2013}. 

In this paper we consider a nanoscale cantilever that consists of a sufficiently large number of atoms to be treated as a classical object. Paramagnetic spins of atomic or nuclear origin, or both, inside the cantilever will be treated as quantum spins flipping due to thermal effects. Damping of micromechanical structures by paramagnetic relaxation in the presence of strong external magnetic field has been studied experimentally  and theoretically in Ref.\ \onlinecite{Harris-APL2003}. It was modeled by the oscillation of the magnetic anisotropy axes of Mn$^{2+}$ ions in the magnetic field due to oscillations of the cantilever. The effect we propose in this paper is not related to the application of the external field or to the nature of the spins. It has its origin in the effective magnetic field generated in the coordinate frame of the cantilever by mechanical rotation. The paper is structured as follows. In Section \ref{Rigid} the physics of the effect is elucidated by considering a rigid oscillating beam that contains paramagnetic spins. Dynamics of a physical elastic cantilever with paramagnetic spins is studied in Section \ref{Elastic}. Discussion of the results and estimates are given in Section \ref{discussion}. 

\section{Rigid beam}

\label{Rigid}

\begin{figure}[ht]
\vspace{-0.15in}
\centering\includegraphics[width=9cm]{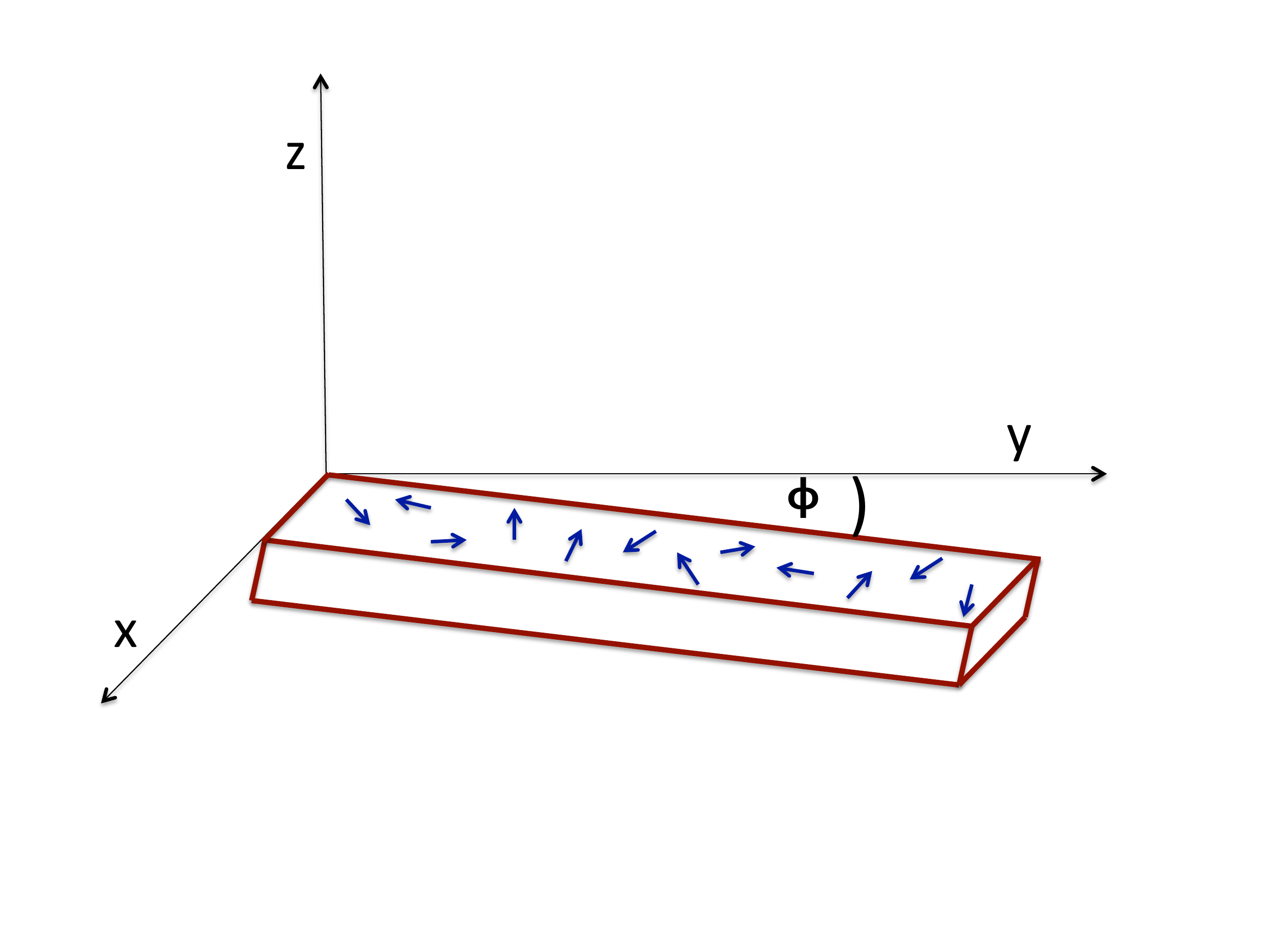} 
\vspace{-0.6in}
\caption{Rigid beam with paramagnetic spins.}\label{rigid}
\end{figure}

Kinematics of a physical cantilever that is shown in Fig. \ref{Cantilever} is more complicated than that of a harmonic oscillator, see, e.g., Ref. \onlinecite{LL-Elasticity}. However, to explain the physics of the effect we will start with a toy model in which the physical cantilever is replaced with a rigid beam that oscillates by changing its orientation with respect to the $y$-axis, with one end to be at the origin of the coordinate frame, see Fig. \ref{rigid}. Its motion is characterized by the angle of rotation, $\phi(t)$, about the $x$-axis. We shall approximate this motion by a harmonic oscillator with a returning torque $\tau_x = -\omega_0^2 \phi$. 
The equation of motion of the beam is ${dJ_x}/{dt}= \tau_x$, 
where $J_x = L_x + S_x$
is the $x$-component of the total angular momentum. The latter consists of the mechanical angular momentum $L_x = I\dot{\phi}$, with $I$ being the moment of inertia, and the spin angular momentum $S_x = \sum_i S^{i}_x$, where the summation is over all spin in the beam. This leads to the following equation of motion
\begin{equation}\label{eq-phi}
I\frac{d^2{\phi}}{d t^2} + I\omega_0^2 \phi = -\hbar \frac{d{S}_x}{dt}
\end{equation}
In most practical situations the mechanical oscillator would be a macroscopic object. It makes sense, therefore, to average the above equation over thermal and quantum fluctuations of the spins,
\begin{equation}\label{eq-phi}
I\frac{d^2{\phi}}{d t^2} + I\omega_0^2 \phi = -\hbar \frac{d}{dt}\langle {S}_x \rangle
\end{equation}

Hamiltonian of the spins, $H_S$, that reflects their interactions in a solid is always written in the coordinate frame that is rigidly coupled to the solid. When the solid rotates the spin Hamiltonian becomes \cite{CGS-PRB2005}
\begin{equation}\label{H2}
H = H_S - \hbar S_x \frac{d \phi}{dt}
\end{equation}
Consequently, the effect of the rotation on the spins is equivalent to the effect of the magnetic field $h = \dot{\phi}/\gamma$, where $\gamma$ is the gyromagnetic ratio for the spin. Thus, one can write
\begin{equation}\label{LR}
\hbar\gamma\langle S_y \rangle = \hat{\chi}\frac{\dot{\phi}}{\gamma}
\end{equation}
where ${\chi}$ is the magnetic susceptibility of the spins.
Switching to Fourrier transforms in equations (\ref{eq-phi}) and (\ref{LR}) one obtains
\begin{equation}\label{F-phi}
I(-\omega^2 + \omega_0^2)\phi_{\omega} = i\hbar \omega \langle S_y\rangle_{\omega}
\end{equation}
\begin{equation}\label{F-S}
\langle {S}_y \rangle_{\omega} = -\frac{i\omega {\chi}(\omega){\phi}_{\omega}}{\hbar \gamma^2}
\end{equation}
Substitution of Eq.\ (\ref{F-phi}) into Eq.\ (\ref{F-S}) then gives
\begin{equation}
\omega^2 = \frac{\omega_0^2}{1 + \frac{\chi(\omega)}{\gamma^2 I}}
\end{equation}
Neglecting renormalization of the real part of the cantilever frequency by the spins and writing $\chi(\omega) = \chi'(\omega_0) + i \chi''(\omega_0)$, $\omega = \omega_0 - i\Gamma$, we get for the rate of damping of the mechanical oscillations
\begin{equation}\label{Gamma}
\Gamma = \frac{\omega_0 \chi''(\omega_0)}{2\gamma^2I}
\end{equation}
where $\chi''$ is the imaginary part of the paramagnetic susceptibility. 

We shall assume that temperature $T$ is high compared to the energy scale of the spin Hamiltonian (\ref{H2}). Then  \cite{MQT-book}
\begin{equation}
\chi''(\omega_0) = f(\omega_0 t_1) \chi_0(T)
\end{equation}
where $\chi_0(T)$ is the equilibrium static Curie susceptibility of $N_S$ quantum spins of length $S$,
\begin{equation}
\chi_0(T) = \frac{N_S\hbar^2 \gamma^2S(S+1)}{3k_B T} 
\end{equation}
and 
\begin{equation}
f(\omega_0 t_1) = \frac{\omega_0 t_1}{1 + (\omega_0 t_1)^2} 
\end{equation}
is the factor depending on the longitudinal spin relaxation time $t_1$.
Substituting this into Eq.\ (\ref{Gamma}) one obtains
\begin{eqnarray}\label{Gamma-rigid}
\Gamma & = & f(\omega_0 t_1)\left[\frac{N_S\hbar S(S+1)}{6I}\right]\frac{\hbar\omega_C}{k_B T} \nonumber \\
&=& f(\omega_0 t_1) \left[\frac{C_S\hbar S(S+1)}{2M_1 L^2}\right]\frac{\hbar\omega_C}{k_B T}
\end{eqnarray}
where we have introduced $I = \frac{1}{3}NM_1L^2$ for the moment of inertia (with $L$ being the length of the beam, $N$ being the number of atoms in the beam, $M_1$ being the mass of one atom) and $C_S = N_S/N$ for the number of spins per atom.   

\section{Physical cantilever}\label{Elastic}

\begin{figure}[ht]
\vspace{-0.6in}
\centering\includegraphics[width=10cm]{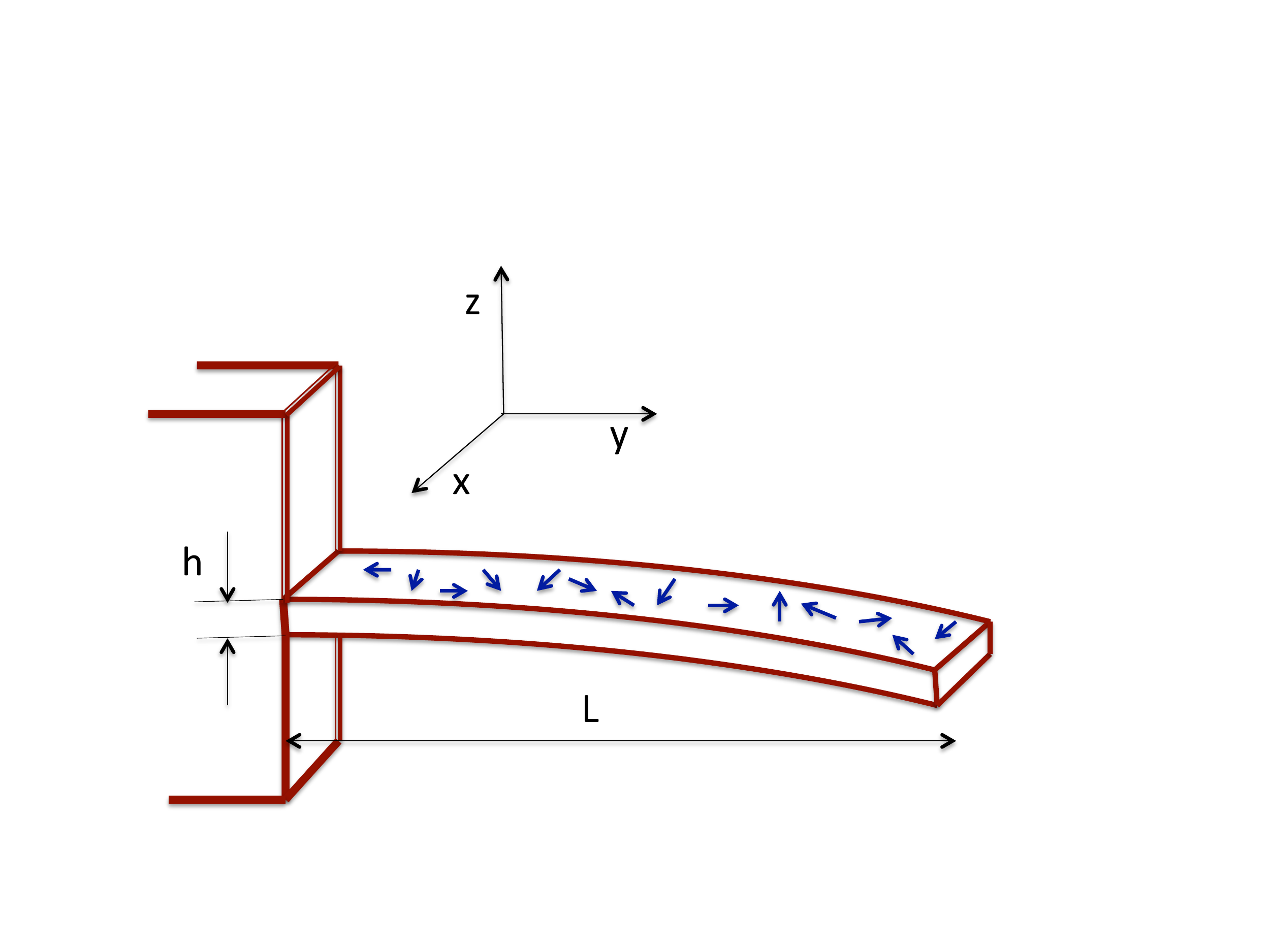} 
\vspace{-0.4in}
\caption{Elastic cantilever with paramagnetic spins.}\label{Cantilever}
\end{figure}

The physical elastic cantilever is shown in Fig. \ref{Cantilever}. Its motion is described by the displacement $u_z(y,t)$ from the equilibrium horizontal position. The dynamical equation for the displacement is \cite{LL-Elasticity}
\begin{equation}\label{elastic}
\rho \frac{\partial^2 u_{\alpha}}{\partial t^2} = \frac{\partial
\sigma_{\alpha \beta}}{\partial x_{\beta}} \,,
\end{equation}
where $\sigma_{\alpha \beta} = {\delta {{H_{tot}}}}/\delta e_{\alpha
\beta}$ is the stress tensor, $e_{\alpha \beta} = \partial
u_{\alpha}/\partial x_{\beta}$ is the strain tensor, $\rho$ is
the mass density of the material, and $H_{tot}$ is the total Hamiltonian of the system. It was shown in Ref. \onlinecite{JCG-PRB2009} that in the presence of the spins the stress tensor can be divided into two parts, the usual elastic part and the part coming from the local internal torques generated by the flipping of the spins. 
The equation that replaces Eq.\ (\ref{eq-phi}) is \cite{JCG-PRB2009}
\begin{equation}\label{elastic-S}
\rho \frac{\partial^2 u_z}{\partial t^2} +
\frac{h^2E}{12(1-\sigma^2)}\frac{\partial^4 u_z}{\partial y^4} =
\frac{\hbar}{2}\frac{\partial}{\partial y}\frac{\partial
}{\partial t}S_{x}(y,t)\,,
\end{equation}
where $\rho$ is the mass density of the cantilever, $h$ is its thickness, $E$ and $\sigma$ are the Young's modulus and the Poisson elastic coefficient ($-1 < \sigma < 1/2$), respectively, and $S_x$ is the $x$-component of the spin density.

Let us write as before $\langle S_x \rangle = \hat{\chi}{\dot{\phi}}/({\hbar \gamma^2})$,
where $\hat{\chi}$ is now the susceptibility of the unit volume. Using the fact that $\phi = {\partial u_z}/{\partial y}$, one has
\begin{equation}
\langle S_x \rangle = \frac{1}{\hbar \gamma^2}\frac{\partial}{\partial t}\hat{\chi}\frac{\partial u_z}{\partial y} 
\end{equation} 
\begin{equation}\label{eq-S}
\rho \frac{\partial^2 u_z}{\partial t^2} +
\frac{h^2E}{12(1-\sigma^2)}\frac{\partial^4 u_z}{\partial y^4} =
\frac{1}{2\gamma^2}\frac{\partial^2}{\partial t^2}\hat{\chi}\frac{\partial^2 u_z}{\partial y^2} 
\end{equation}
It is convenient to switch to dimensionless variables,
\begin{equation}\label{bar}
\bar{u}_z = \frac{u_z}{L}\,, \quad \bar{y} = \frac{y}{L}\,, \quad
\bar{t} = t\nu\,, \quad \nu \equiv
\sqrt{\frac{Eh^2}{12\rho(1-\sigma^2)L^4}}\,,
\end{equation}
where $\nu$ determines the scale of the eigenfrequencies of the cantilever. In terms of these variables Eq.\ (\ref{eq-S}) becomes
\begin{equation}\label{eq-bar}
\frac{\partial^2 \bar{u}_z}{\partial \bar{t}^2} + \frac{\partial^4
\bar{u}_z}{\partial \bar{y}^4} = \frac{1}{2\gamma^2\rho L^2}\frac{\partial^2}{\partial \bar{t}^2}\hat{\chi}\frac{\partial^2 \bar{u}_z}{\partial \bar{y}^2} 
\end{equation}
This equation has to be solved with the boundary conditions $\bar{u}_z = 0$, ${\partial \bar{u}_z}/{\partial \bar{y}}
= 0$ at $\bar{y} = 0$ and ${\partial^2 \bar{u}_z}/{\partial \bar{y}^2} = 0$, ${\partial^3 \bar{u}_z}/{\partial \bar{y}^3} = 0$ at $\bar{y} = 1$.
The first two conditions correspond to the absence of the displacement
and the absence of the bending of the cantilever at the fixed end,
while the last two conditions correspond to the absence of the torque
and the force, respectively, at the free end.

For the free oscillations of the cantilever in the absence of the spins one
writes
\begin{equation}\label{free}
\bar{u}_z(\bar{y}, \bar{t}) =
\bar{u}(\bar{y})\cos(\bar{\omega}\bar{t})\,.
\end{equation}
Substitution into Eq.\ (\ref{eq-bar}) with $\hat{\chi} = 0$ then
gives
\begin{equation}\label{eq-u}
\frac{\partial^4 \bar{u}}{\partial \bar{y}^4} - \kappa^4\bar{u} =
0\,, \qquad \kappa^2 \equiv \bar{\omega}\,.
\end{equation}
Solution of this equation with the boundary conditions gives \cite{LL-Elasticity}
\begin{eqnarray}\label{eigenfunction}
& & \bar{u}(\bar{y}) = (\cos\kappa + \cosh\kappa)\left[\cos(\kappa
\bar{y}) - \cosh(\kappa
\bar{y})\right] \nonumber \\
& & + (\sin\kappa - \sinh\kappa)\left[\sin(\kappa \bar{y})-
\sinh(\kappa \bar{y})\right]
\end{eqnarray}
with
\begin{equation}\label{modes}
\cos\kappa\cosh\kappa + 1 = 0\,,
\end{equation}
for the frequencies of the normal modes of the cantilever,
$\bar{\omega}_n = \kappa^2_n$ (measured in the units of $\nu$ of
Eq.\ (\ref{bar})). Fundamental (minimal) frequency is
$\bar{\omega}_1 \approx 3.516$. The next two frequencies are
$\bar{\omega}_2 \approx 22.03$ and $\bar{\omega}_3 \approx 61.70$. 
The profiles of the oscillations of the cantilever for the first three
normal modes ($n=1,2,3$) are shown in Fig. \ref{modes}.
\begin{figure}[ht]
\begin{center}
\vspace{-0.1in}
\includegraphics[width=65mm, angle = -90]{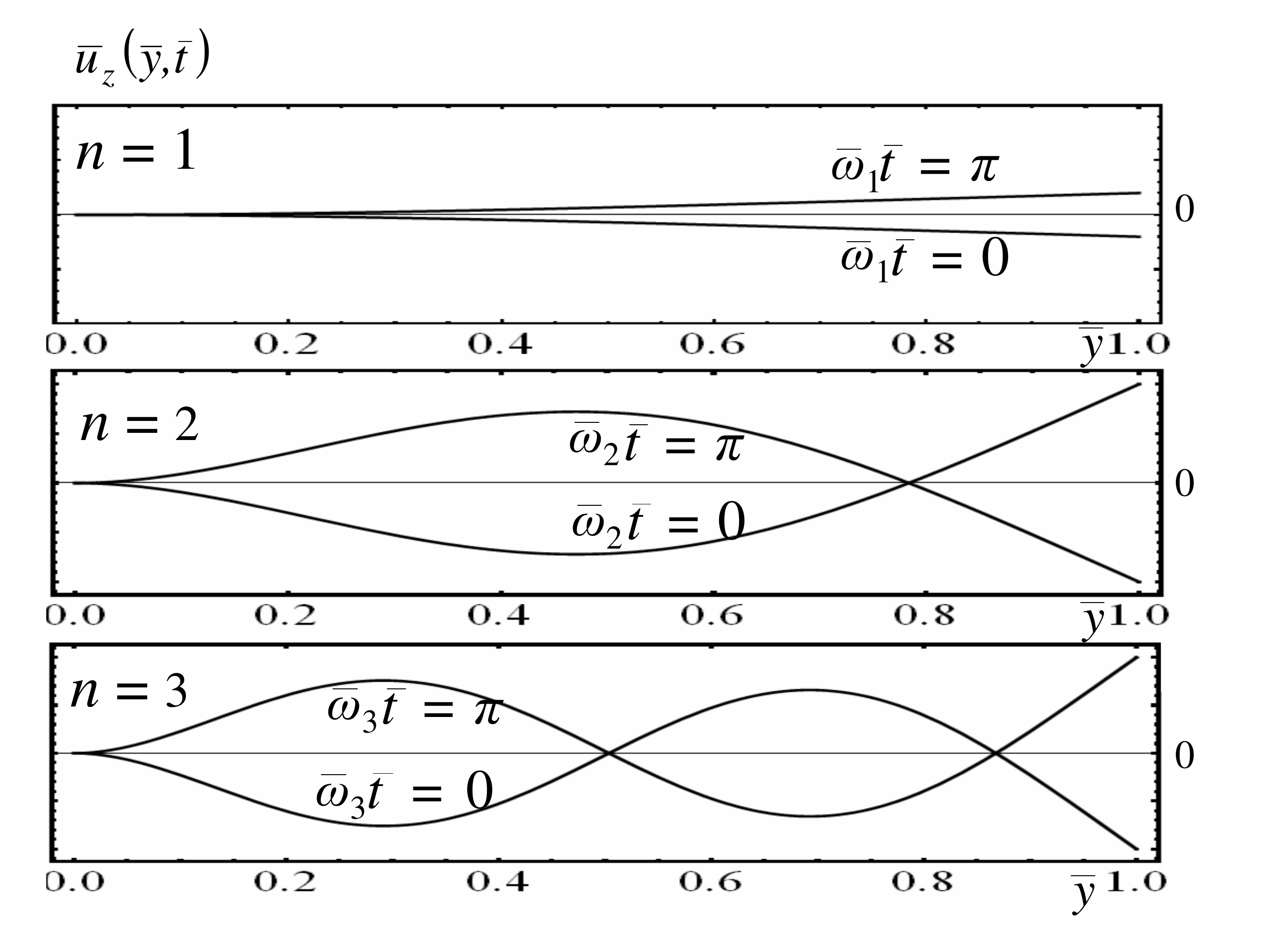}
\caption{Profiles of the oscillating cantilever at different
moments of time for $n = 1,2,3$. } \label{modes}
\end{center}
\end{figure}

With account of the term in the right hand side of Eq.\ (\ref{eq-bar}) equation (\ref{eigenfunction}) becomes
\begin{equation}\label{eq-u}
\frac{\partial^4 \bar{u}}{\partial \bar{y}^4} - \kappa'^4\bar{u} = -\frac{\kappa'^4}{2\gamma^2\rho L^2}{\chi}_{\omega}\frac{\partial^2 \bar{u}}{\partial \bar{y}^2} 
\end{equation}
Since the right hand side of this equation is small, to obtain the frequency $\bar{\omega}'^2 = \kappa'^4$ renormalized by the presence of the spins, one can safely substitute here the eigenmode of Eq.\ (\ref{eigenfunction}), for which ${\partial^2 \bar{u}}/{\partial \bar{y}^2} = - \kappa^2 \bar{u}$,  ${\partial^4 \bar{u}}/{\partial \bar{y}^4} = \kappa^4 \bar{u}$. This gives
\begin{equation}
{\omega'}_n^2 = \frac{\omega_n^2}{1 + \frac{\kappa_n^2\chi(\omega_n)}{2\gamma^2\rho L^2}}
\end{equation}
The imaginary part of the frequency is
\begin{equation}
\Gamma_n = \frac{\omega_n \kappa_n^2}{2\gamma_n^2\rho L^2}\chi''(\omega_n)
\end{equation}
with
\begin{equation}
\chi''(\omega_n) = f(\omega_n t_1)\left[\frac{n_S\hbar^2 \gamma^2S(S+1)}{3k_B T} \right]
\end{equation}
where $n_s = N_S/V = C_S (\rho/M_1)$ is the number of spin per unit volume. Consequently
\begin{equation}\label{Gamma-bending}
\Gamma_n =f(\omega_n t_1)\left[\frac{C_S \kappa_n^2\hbar S(S+1)}{6M_1 L^2}\right]\frac{\hbar\omega_n}{k_B T}
\end{equation}

For the first eigenmode, $\kappa_1^2 \approx 3.516$, the damping rate of the physical cantilever, given by Eq. (\ref{Gamma-bending}), is greater than the damping rate of the rigid harmonic beam, given by Eq.\ (\ref{Gamma-rigid}), by a factor $3.516/3 = 1.172$. Notice, however, that the corresponding factor becomes significantly greater for higher modes, $\kappa_2^2/3 \approx 7.343$, $\kappa_3^2/3 \approx 20.57$ and so on. This is because $d{\phi}/dt$ and the corresponding effective magnetic field acting on the spins in the rotating frame, $\dot{\phi}/\gamma$, is greater for higher modes, see Fig. \ref{modes}.\\

\section{Discussion}\label{discussion}

We have computed the contribution of paramagnetic spins to the damping of the mechanical oscillations of the cantilever. Eq.\ (\ref{Gamma-bending}) provides the damping rate of the $n$-th mode in terms of the concentration, $C_S$, of spins of length $S$, flipping with the time constant $t_1$. Since $\Gamma_n$ depends on parameters that are usually known in experiment, it can be easily estimated for a given cantilever. When different kinds of spins are present, they contribute to the damping additively in accordance with Eq.\ (\ref{Gamma-bending}).  Note that $f(\omega_n t_1)$ has a maximum at $\omega_n t_1 = 1$. Thus, at comparable concentrations, the spins that flip at a rate comparable to $\omega_n$ provide the maximal damping. For $\omega_n$ in the kHz range these would normally be the nuclear spins, while for $\omega_n$ in the GHz range these would be the atomic spins. 

The physical mechanism of the damping studied in this paper is this. The effective ac magnetic field on the spins produced by the oscillations of the cantilever originates from the non-inertial effect of the local crystal fields \cite{Hehl-PRD1990,CGS-PRB2005}. It couples the dynamics of the cantilever with the dynamics of the spins. The latter is constantly disturbed by thermal phonons that make the spins flip.  The resulting local torques in the crystal lattice are mandated by the conservation of the angular momentum. They transfer angular momentum to the cantilever, causing damping. It is the spin-phonon interaction that is responsible for the damping and for the conversion of the mechanical kinetic energy of the cantilever into its thermal energy. 

The quality factor of the cantilever is $Q_n = \omega_n/\Gamma_n$. In practical situations one would want to know if the quality factor observed in experiment had anything to do with the spins. To answer this questions we notice that the maximal value of $f$ is $1/2$. Consequently, the spins cannot make the quality factor lower than
\begin{equation}
Q_{min}^{(n)} = \frac{12 M_1 L^2 k_BT}{\hbar^2 k_n^2 C_S S(S+1)}
\end{equation}
In the kelvin temperature range the value of $Q_{min}^{(n)}$ due to spins for a small cantilever of length $L \sim 10$nm and high concentration of paramagnetic spins can be of order of $10^3$  at $n = 1$ and progressively lower at higher $n$. This suggests that paramagnetic spins should be suspect when the quality factor of a nanoscale cantilever becomes small on decreasing temperature. This mechanism of damping can also apply to the oscillations of nanowires and macromolecules. 

\section{Acknowledgements}
This work has been supported by the U.S. National Science Foundation through Grant No. DMR-1161571.

\end{document}